\documentclass[12pt,oneside]{article} 

\usepackage{amssymb,amsmath,amsthm,enumerate,hyperref,mathtools}
\usepackage{graphics,latexsym,amsfonts}  
\usepackage[dvipsnames]{xcolor} 
\usepackage{ulem}  
\usepackage{mathrsfs} 
\hypersetup{
    colorlinks, 
    linkcolor={red!50!black}, 
    citecolor={blue!50!black},
    urlcolor={blue!80!black} 
}

\usepackage[displaymath, mathlines]{lineno}  
\usepackage{mathabx}
\usepackage{graphicx,latexsym,amsfonts}   
\usepackage{lineno}
\usepackage{picture}
\usepackage{picture}
\usepackage[english]{babel} 
\usepackage{enumitem}  
\usepackage{comment}
\setlist[enumerate]{leftmargin=*}    
\usepackage{color,soul}
\usepackage[longnamesfirst]{natbib}

\def\vs{\vspace{-0.1cm}}
\renewcommand{\emph}{\textsl}

\def\reference#1{\href{#1}{Cliquer ici pour voir une r\'ef\'erence.}}

\usepackage[longnamesfirst]{natbib}

\usepackage{setspace}
\usepackage[top=1.25in,bottom=1.25in,left=1.25in]{geometry}
\def\NN{\mathbb{N}} 

\def\QQ{\mathbb{Q}}

\def\U{{\mathcal U}}
\def\W{{\mathcal W}}

\def\defbyarrow{\stackrel{\mathrm{def}}{\Longleftrightarrow}}

\def\es{\varnothing}

\newtheorem*{theorem*}{Theorem}
\newtheorem*{claim*}{Claim}
\newtheorem*{problem*}{Problem}

\newtheorem{theorem}{Theorem}
\newtheorem{corollary}{Corollary}
\newtheorem{proposition}{Proposition}

\newtheorem{remark}{Remark}

\newtheorem{lemma}{Lemma}
\newtheorem{example}{Example}
\newtheorem{definition}{Definition}

\def\dem{\textsf{dem}}
\def\lib{\textsf{lib}}

\def\es{\varnothing}

\mathchardef\ordinarycolon\mathcode`\:
\mathcode`\:=\string"8000
\begingroup \catcode`\:=\active
  \gdef:{\mathrel{\mathop\ordinarycolon}}
\endgroup

\title{Rationalization of indecisive choice behavior by majoritarian ballots\footnote{The authors are grateful to Davide Carpentiere for several insightful comments. Jos\'e Carlos R.\ Alcantud thanks the audience of The Saarland Workshop in Economic Theory for their suggestions. Alfio Giarlotta gratefully acknowledges the financial support of ``Ministero dell'Istruzione, dell'Universit\`a e della Ricerca (MIUR) - PRIN 2017'', project \textit{Multiple Criteria Decision Analysis and Multiple Criteria Decision Theory}, grant 2017CY2NCA. }} 

\author{\normalsize Jos\'{e} Carlos\ R.\ Alcantud\thanks{BORDA Research Unit and IME, University of Salamanca, Spain. jcr@usal.es}$\:$,
Domenico Cantone\thanks{Department of Mathematics and Computer Science, 
University of Catania, Italy. domenico.cantone@unict.it}$\:$,
Alfio Giarlotta\thanks{Department of Economics and Business,
University of Catania, Italy. alfio.giarlotta@unict.it \textit{(corresponding author)}}$\:$,
Stephen Watson\thanks{Department of Mathematics and Statistics, York University, Toronto, Canada. swatson@yorku.ca}}

\begin{document}

\makeatletter

\def\@fnsymbol#1{\ensuremath{\ifcase#1\or a\or b\or c \or
   d \or e\or \| \or 7\or 8 \or 9 \else\@ctrerr\fi}}
   
\maketitle
\makeatother


\begin{abstract} 
\noindent We describe a model that explains possibly indecisive choice behavior, that is, quasi-choices (choice correspondences that may be empty on some menus). 
The justification is here provided by a proportion of ballots, which are quasi-choices rationalizable by an arbitrary binary relation. 
We call a quasi-choice $s$-majoritarian if all options selected from a menu are endorsed by a share of ballots larger than $s$.
We prove that all forms of majoritarianism are equivalent to a well-known behavioral property, namely Chernoff axiom. 
Then we focus on two paradigms of majoritarianism, whereby either a simple majority of ballots justifies a quasi-choice, or the endorsement by a single ballot suffices -- a liberal justification.  
These benchmark explanations typically require a different minimum number of ballots.  
We determine the asymptotic minimum size of a liberal justification.  
 
\medskip


\medskip


\end{abstract} 


\section{Introduction} \label{SECT:intro}

%

Many models of multi-self decision making have been designed with the purpose of explaining context-dependent behaviour: see, e.g., \cite{AmbrusRozen2014}, \cite{FudenbergLevine2006}, \cite{ManziniMariotti2007}, \cite{May1954}, \cite{SilvaLeanderSeth2017}.
In a variety of manners, they accommodate choices that violate the property of \textsl{Independence of Irrelevant Alternatives} \textsl{(IIA)}.
Some authors \citep{AmbrusRozen2014} have even measured the degree of irrationality of a choice function by the number of violations of \textsl{IIA} that is shows.
In this paper we study a model that does the opposite job.
Our observable is a possibly indecisive choice correspondence, i.e., an utmost general expression of selections. 
Then within a multi-selves formulation, the model explains exactly all choices that satisfy \textsl{IIA}.

For comparison, consider the model of bounded rationality (\textsl{rationalization by multiple rationales}, \textsl{RMR}) elaborated by \cite{KalaiRubinsteinSpiegler2002}.
The authors argue that violations of principles of rational selection can be explained if every choice set has an epistemic value, and the decision maker (DM) applies a suitable rationale for the specific choice problem posed by that menu.\footnote{Compare to \cite{Sen1993}, who provides arguments for `rational' ways to contradict the property of \textsl{IIA}.}
The choice set triggers the application of one of the feasible rationales (linear orderings), and the selection may be dependent on the information that it embodies. 

Since the RMR model is non-testable -- that is, all choices do admit a multiple rationalization -- its focus is not on rationality \textit{per se}, but instead on `degrees of rationality': in fact, in Section~4 of the mentioned paper, the authors say that ``\textit{the appeal of the explanation depends on the number of orderings involved}''.
Probably because choice functions  are single-valued (a unique item is selected from each menu), in the RMR model the epistemic value of the chosen item is disregarded, or at least amalgamated with that of the menu.
As soon as the DM knows the choice problem, she selects the rationale that produces the desired result, irrespective of the information that this choice conveys.\footnote{
Remark \ref{REM:KRS} below refers to the application of the RMR spirit to choice correspondences.
For a structured version of the RMR model, where the justifying rationale is linked to the most `salient' items of the menu itself, see \citet{GiaPetrWat2022}.}

Suppose, on the other hand, that the DM is allowed to select more than one item from a menu -- that is, we deal with a multi-valued choice, also called a choice correspondence. 
Then it may be the case that she finds different reasons for testing the suitability of each of her selections.
To illustrate this point, consider the example used by \cite{EliazOk2006} to argue how incomplete preferences can explain violations of the \textsl{Weak Axiom of Revealed Preference (WARP)}.  

\begin{example} \label{EX:Watson} (Mrs.\ Watson's movie selection) \rm 
(\citealp{EliazOk2006}, Example~1)
Mrs.\ Watson has to choose a movie to rent for her children, Alice and Tom.    
The set of available movies is $X=\{x, y, z\}$. 
Alice's and Tom's preferences over $X$ are represented by the following rankings: 
$y \succ _A z \succ _A x$ (for Alice), and $x \succ _T y \succ _T z$ (for Tom).
Then Mrs.\ Watson's choice $c$ is as follows (selected items are underlined):\footnote{This definition seems redundant, because choices from menus with a single item are forced in case at least an element must be selected from every nonempty menu. However, in this paper we also consider indecisive DMs, whose selections are quasi-choices, that is, selecting no item at all from a menu becomes a feasible option.}\vs\vs 
\begin{equation*} \label{EQ:Watson}
	\underline{x}\underline{y}z\,, \; \underline{x}\underline{y}\,,\; \underline{x}\underline{z}\,,\; \underline{y}z\,,\; \underline{x}\,,\;\underline{y}\,,\;\underline{z}\,,
\end{equation*}
where a decision like $c(\{x,z\}) = \{x, z\}$ (denoted by $\underline{x}\underline{z}$ for brevity) means that she will flip a coin to decide which movie to rent between $x$ and $z$. 
Although Mrs.\ Watson's approach contradicts WARP (because $z$ is chosen in both $\underline{x}\underline{z}$ and $\underline{y}\underline{z}$ but not in $\underline{x}\underline{y}z$), \cite{EliazOk2006} illustrate how plausible her way to settle conflicts may be considered. 

Alternatively -- and more in the flavor of this paper -- Mrs.\ Watson's selections might be explained by the following mental process.
In any choice problem (say, the selection from the menu $\{y,z\}$), she flips a coin between the eligible movies that are \textit{best for at least one of the children.}
Then her apparently conflicting choices become a natural consequence of her attitude.
Indeed, she explains her choice from the menu $\{x,z\}$ by the fact that both Tom's favorite ($x$) and Alice's favorite ($z$) cannot be disregarded. 
For a similar reason, $z$ is rejected from $\{x,y,z\}$, because it is never the best in any of the two available rankings.  
Mrs.\ Watson's selections are all justified by this mechanism: the flip of a coin between the best alternatives for at least one of the two rationales, namely Alice's and Tom's preferences.
\end{example}

The latter argument is subtly different from the spirit of the RMR model in a critical feature: Mrs. Watson is (voluntarily) bound to the opinions of her children.
When the rationales are fixed, her opinion is determined by them.
The rationales are not a mere consultative tool, instead they enforce actions.
To emphasize the role of the rationales in the process of rationalization, we shall refer to them as `voters', and the selections they enforce will be called `ballots'.
Our approach departs from the RMR model of \cite{KalaiRubinsteinSpiegler2002} in three main aspects.\vs
\begin{description}
\item[\rm (1)]
The items selected from each menu may not be unique, and there might be empty choices.
The range of our analysis is much wider, because our observables are quasi-choice correspondences rather than choice functions.
In other words, from any menu $A$ it is possible to select no item at all; furthermore, when we observe the selection of $x$ from $A$, we are only allowed to infer $x\in c(A)$, and not $c(A) = \{x\}$.\vs  
\item[\rm (2)]
We do not ignore the epistemic value of the alternatives.
The DM can find different reasons for a selection: for each menu, she ponders whether a justification exists for choosing any given item.\vs
\item[\rm (3)]
Justifications are not interpreted as in the RMR model, which permits the existence of rationales that are never applied.
On the contrary, we do not allow irrelevant opinions: a justification `is' a voter, who may select all, some, or none of the options available in a menu via the ballot (that is, the quasi-choice) derived from her opinion. 
In other words, justifications stem from a natural voting procedure involving all justifications.
\end{description}

Another feature of our approach is that it makes simultaneous use of the justifications.
This characteristic distinguishes our approach from models where rationales are sequentially applied,
e.g., the \textit{Rational Shortlist Methods} \textit{(RSM)} in \cite{ManziniMariotti2007}.

With these minor modifications, we can draw a neat demarcation line between `rational' and `non-rational' choice behavior.   
In fact, the explanation of a choice correspondence by means of a voting process in the sense hinted at above gives rise to a testable model, as the next example shows.  

\begin{example} \textit{(A non-liberal choice)} \rm
	Suppose that a DM chooses from a grand set $X = \{x,y,z\}$ as follows: $c(\{x,z\}) = \{x\}$, and $c(A) = A$ for any other menu. 
    According to the most basic presentation of our model, a \textsl{liberal} justification of this choice behavior is a finite collection $V$ of voters over $X$ such that the options selected from each menu are exactly those chosen by\textit{ at least one} voter in $V$.   
    Here we represent a \textsl{voter} by an unrestricted binary relation $\to_i$ on $X$, where `unrestricted' means that $\to_i$ is an arbitrary subset of $X \times X$, possibly displaying loops or cycles.  
    In other words, a liberal justification of $c$ consists of a finite family $\{\to _i \; : \: i \in I\}$ of binary relations over  $X$ such that, for each menu $A \subseteq X$ and candidate $p \in A$, we have\vs\vs
\begin{equation*} \label{EQ:preliminary} 
		p \in c(A) \quad \Longleftrightarrow \quad p \in \max(A,\to_i) \mbox{ for some } i \in I.\vs\vs 
\end{equation*}
(As usual, $\max(A,\to_i)$ is the subset of $A$ composed of the items $p$ for which there is no $q \in A$ such that $q \to_i p$ holds.) 
Note that such a binary relation $\to_i$ over $X$ naturally induces a quasi-choice $v_i$ over $X$, defined by $v_i(A) = \max(A,\to_i)$ for all $A \subseteq X$: we call such a quasi-choice the \textsl{ballot} over $X$ derived from $\to_i$. 
(In technical terms, a ballot over $X$ is a rationalizable quasi-choice.) 
Therefore, a liberal justification for $c$ is equivalently given by a family $\{v_i : i \in I\}$ of ballots over $X$ such that for each $A \subseteq X$ and $p \in A$, we have\vs\vs 
\begin{equation*} \label{EQ:preliminary_bis} 
		p \in c(A) \quad \Longleftrightarrow \quad p \in v_i(A) \mbox{ for some } i \in I.\vs\vs 
\end{equation*}
It turns out that such a family (of voters, or, equivalently, ballots) does not exists, and so $c$ cannot be explained by the liberal paradigm.
We conclude that this choice behavior must be labeled \textsl{non-rational} by the simplest instance of our approach.
\end{example}

A direct proof of the non-liberality of $c$ is not immediate, despite the tiny number of alternatives. 
A simpler proof relies on the normative implications of this model: in fact, liberality is equivalent to a classical condition of choice consistency, namely \cite{Chernoff1954} property, also called Axiom$\:\alpha$ by \cite{Sen1971}.\footnote{This fact is known: see \cite{AizermanAleskerov1995}.} 
Then $c$ cannot be explained by the liberal case of our model, because Axiom$\:\alpha$ is violated: $z$ is chosen in the grand menu $X$ but is not chosen in its submenu $\{x,z\}$. 	


Arguably, the liberal model described above appears to be excessively permissive, as a single rationale (equivalently, a single ballot) suffices to ensure the election of a candidate.  
In contrast, consider a different version -- here called democratic -- of our approach, in which an option is selected from a menu if and only if it is chosen \textit{by more than half} of the voters/ballots.
More formally, a quasi-choice correspondence $c$ over $X$ is \textsl{democratic} if there is a finite family $\{\to_i : i \in I\}$ of voters over $X$ such that, for each $A \subseteq X$ and $p \in A$, we have\vs\vs
\begin{equation*} \label{EQ:preliminary2}
		p \in c(A) \quad \Longleftrightarrow \quad p \in \max(A,\to_i) \mbox{ for more than } \textstyle \frac{\vert I \vert}{2}\mbox{ voters}.\vs\vs	
\end{equation*}
Alternatively, $c$ is democratic, if there is a finite family $\{v_i : i \in I\}$ of ballots over $X$ such that, for each $A \subseteq X$ and $p \in A$, we have\vs\vs
\begin{equation*} \label{EQ:preliminary2_bis}
		p \in c(A) \quad \Longleftrightarrow \quad p \in v_i(A) \mbox{ for more than } \textstyle \frac{\vert I \vert}{2}\mbox{ ballots}.\vs\vs	
\end{equation*}
Such a justification appears to have more solid grounds than a liberal one, because a democratic support for a decision hints at a robust behavior.

However -- and maybe surprisingly -- the normative implication of a democratic model is exactly the same as that of the liberal one: in fact, we shall show that a quasi-choice is democratic if and only if Chernoff property holds. 
As a consequence, outside observers who only monitor the choices made by a DM cannot discard that she has taken the `democratic' position, once they have confirmed that the `liberal' mechanism justifies her choices. 

In fact, more is true: any majoritarian modification of the RMR model is equivalent to Axiom$\;\alpha$, where by `any' we mean that the share of voters that witnesses the selection of an item may be any number between zero and one. 
More formally, given a `share' $s \in [0,1)$, we say that a quasi-choice correspondence $c$ over $X$ is \textit{$s$-majoritarian} if there exists a finite family $\{\to _i \; : \: i \in I\}$ of voters over $X$ such that, for each menu $A \subseteq X$ and alternative $p \in A$, we have\vs\vs 
\begin{equation*} \label{EQ:preliminary3}
		p \in c(A) \quad \Longleftrightarrow \quad p \in \max(A,\to_i)  \mbox{ for more than } s\vert I \vert \mbox{ voters.}\vs\vs	
\end{equation*}
(Again, an equivalent version of this property can be given in terms of ballots.) 
According to this general notion, liberal means $0$-majoritarian, and democratic means $0.5$-majoritarian. 
In this paper, we prove that being $s$-majoritarian for any share $s \in [0,1)$ is equivalent to the satisfaction of Chernoff property: see \textit{Theorem~\ref{THM:main1}} and \textit{Corollary~\ref{COR:full_generality}}.
This result provides a full generalization of the characterization given by \cite{AizermanAleskerov1995}, whereby liberalism is extended to all types of majoritarianism. 

Now a natural question arises: \textit{How can we distinguish a specific majoritarian paradigm from a different one?}
We shall partially answer this query by focusing on liberalism and democracy. 
Similarly to the RMR model -- where the minimum number of linear rationales justifying a single-valued choice yields a  `degree of rationality' of that behavior -- we distinguish liberal from democratic behavior by the minimum number of voters that provide a rationalization. 
Such proxies are respectively called the liberal and democratic numbers of the observed choice correspondence.   
Specifically, we examine the mutual relationship between the `liberal number' and the `democratic number' as the size of the grand set of candidates varies. 
Our analysis yields a tight upper bound for the liberal number of a choice satisfying Axiom$\;\alpha$ (\textit{Theorem~\ref{THM:upper bound lib}}).
Then, using Stirling's approximation, we determine the asymptotic behavior of the minimal size of a liberal justification (\textit{Corollary~\ref{COR:limit behavior of lib}}).   

\smallskip

The paper is organized as follows. 
Section~\ref{SECT:preliminaries} collects preliminaries. 
In Section~\ref{SECT:multi-rationalizations_by_voters} we give the notion of rationalization by ballots, and characterize it by Chernoff property.  
In Section~\ref{SECT:numbers} we distinguish two specifications of our model, liberalism and democracy, and determine a tight upper bound for the minimum number of voters witnessing liberalism.  
Section \ref{SECT:conclusions} concludes the paper. 
Minor proofs are presented in the main body of the paper, whereas main proofs are in Appendix.


\section{Rationalization, ballots, and voters} \label{SECT:preliminaries}

Let $X$ be a finite set of $n \geqslant 2$ alternatives (also called items or candidates); we refer to $X$ as the \textsl{grand set}.  
As usual, $2^X$ denotes the family of all subsets of $X$.  
A \textsl{quasi-choice correspondence} over $X$ is a map $c \colon 2^X \to 2^X$  that satisfies\vs  
\begin{description}
	\item[\rm \textsf{(contractiveness)}] $c(A) \subseteq A$ for each $A \in 2^X$.\vs
\end{description}
If, in addition, $c$ satisfies\vs 
\begin{description}
  \item[\rm \textsf{(decisiveness)}] $c(A) \neq \es$ when $A \neq \es$,\vs
\end{description} 
then it is a \textsl{choice correspondence}. 
Historically, only choice correspondences have been an object of careful study, and quasi-choices correspondence have been disregarded.   
However, very recently \cite{Costa-Gomes2022} have reported on choice experiments that strongly suggest that nonforced-choice models rejecting decisiveness may offer a powerful lens to study revealed preferences.
Following this stream of evidence, our main results avoid the assumption of decisiveness, and concern quasi-choice correspondences in general.

Hereafter, unless ambiguity may arise, we shall drop the term `correspondences', and refer to choice correspondences and quasi-choice correspondences as \textsl{choices} and \textsl{quasi-choices}, respectively. 
A very special case of choice is given by a single-valued choice $c$, that is, $\vert c(A) \vert = 1$ for each nonempty $A \in 2^X$, which is classically referred to as a \textsl{choice function}.
In this case, $c$ is identified with the function $c \colon 2^X \setminus \{\es\} \to X$ defined in the obvious way. 

Each $A \in 2^X$ is a \textsl{menu}, and $c(A)$  is the \textsl{choice set} of $A$, which collects all the items in $A$ that are deemed selectable by the DM.  
If $c$ is a choice, then at least one item is selected from each nonempty menu.
Instead, for a quasi-choice the DM is allowed `not to decide', selecting no item from some (even all) menus.  
The two extreme types of quasi-choices over $X$ are therefore the following:
\vs 
\begin{itemize} 
	\item the \textsl{identity choice} -- a `neutral judgement' -- in which every candidate is always selected from any menu, that is, $c(A) =A$ for all $A \in 2^X$;\vs
	\item the \textsl{null quasi-choice} -- a `hypercritical judgement' -- in which no candidate is selected from any menu, that is, $c(A) = \es$ for all $A \in 2^X$.\vs 
\end{itemize}
In our setting, the identity choice can be thought of as representative of a mild and socially responsible behavior, when there is no substantial argument to reject any of the feasible candidates. 
On the contrary, a null quasi-choice encodes a behavior that either abhors all available candidates or simply wants no (even indirect) involvement in any political matter.  

A binary relation $\succ$ over $X$ is an arbitrary subset of $X \times X$. 
In particular, $\succ$ is \textsl{asymmetric} if $x \succ y$ implies $\neg(y \succ x)$ for all $x,y \in X$; in this case, $\succ$ is also \textsl{loopless} (or \textsl{irreflexive}), that is, $x \succ x$ holds for no $x \in X$.  
The classical notion of rationality in \textsl{revealed preference theory} is encoded by the existence of an asymmetric relation that explains choice behavior by maximization \citep{Samuelson1938}:   
  
\begin{definition} \rm \label{DEF:asymmetric_rationalizability}
A quasi-choice $c$ over $X$ is \textsl{asymmetrically rationalizable}
if there is an asymmetric relation $\succ$ over $X$ such that $c(A) = \max (A, \succ~\!\!)$ for all $A \in 2^X$, where $\max(A, \succ)$ is the set $\{a \in A : b \succ a \hbox{ for no } b \in A\}$. 
\end{definition}

Since we interpret $b \succ a$ as `candidate $b$ dominates candidate $a$', the set $\max(A, \succ)$ collects all non-dominated candidates of $A$.  
Note that if $c$ is an asymmetrically rationalizable choice, then the asymmetric relation $\succ$ that rationalizes $c$ is unique; moreover, $\succ$ must also be \textsl{acyclic}, that is, there is no \textsl{cycle} $x_1 \succ x_2 \succ \ldots \succ x_k \succ x_1$ of $k \geqslant 3$ items. 

The notion of rationality given in Definition~\ref{DEF:asymmetric_rationalizability} does not fit well the very concept encoded by a quasi-choice, because it may fail to consistently explain indecisive choice behavior.  
Indeed, if $c$ is a quasi-choice that is asymmetrically rationalizable, then all menus $A$ of size one or two \textit{must} have a nonempty choice set $c(A)$ by the very definition of asymmetric rationalizability.  
In other words, a rational DM is allowed not to decide \textit{only} for menus with at least three items, whereas she is forced to select at least one item from all menus of size two and one. 
The possibility of an empty selection is banned for menus of the type $\{x,y\}$ or $\{x\}$. 
Is such a constraint justifiable?

In this regard, \cite{Costa-Gomes2022} report on experimental findings in psychology with a story from~\cite{Shafir1993}: Thomas Schelling ``\textit{was presented with two attractive encyclopedias and, finding it difficult to choose between the two, ended up buying neither [...]  Had only one encyclopedia been available he would have happily bought it.}''
This phenomenon has been demonstrated in experiments with one and two alternatives~\cite[Section 4]{TverskyShafir1992}. 
To give a simple example, in a political ballot, if there are only two candidates $x$ and $y$ who belong to extreme wings, and I have very moderate political views, then I may decide not to vote at all (that is, $c(\{x,y\}) = \es$). 
By the same token, if I am allergic to chocolate, and a restaurant only offers a chocolate cake $x$ as dessert, then I will simply avoid taking dessert (that is, $c(\{x\}) = \es$). 

The awkward restriction on indecisive choices described above motivates the next definition, where the notion of rationality is relaxed, in order to model the very concept of an indecisive DM.
 
\begin{definition} \rm \label{DEF:free_rationalizability}
  A quasi-choice $c$ over $X$ is \textsl{freely rationalizable} (\textsl{rationalizable}, in case no ambiguity may arise) if there is a binary relation $\to$ over $X$ such that\vs\vs 
\begin{equation*}\label{representability} 
  c(A) = \max (A,\to) = \{a \in A \, : \, b \not\rightarrow a \hbox{ for all } b \in A\}\vs\vs
\end{equation*}
  for each $A \in 2^X$, where $b  \not\rightarrow a$ stands for the negation of $b \to a$. 
  Hereafter, any freely rationalizable quasi-choice on $X$ is called a \textit{ballot} over $X$, and is often denoted by $v,w,z$, etc. (instead of $c$). 
When no confusion may arise, we shall abuse notation, and identify a ballot over $X$ with any binary relation $\to$ over $X$ that rationalizes it: we call this binary relation a \emph{voter}.\footnote{Note that the binary relations (voters) that rationalize a ballot need not be unique. However, this abuse of notation will be harmless.} 
\end{definition}

If a quasi-choice $c$ is asymmetrically rationalizable, then we use $\succ$ for the (unique) rationalizing preference to avoid confusion. 
Instead, for a ballot $v$, we use the `arrow notation' $\to$ for any rationalizing preference (voter). 
Note that, in the latter case, the rationalizing relation $\to$ may\vs
\begin{itemize}
  \item[(a)] display loops, that is, $x \to x$ holds for some $x \in X$, in which case we call the item $x$ \textsl{repellent}, and/or\vs 
  \item[(b)] fail to be asymmetric, hence $x \to y$ and $y \to x$ are allowed to coexist for distinct $x,y \in X$, in which case we call the items $x,y$ \textsl{pairwise exclusive}, and use the notation $x \leftrightarrows y$.
\end{itemize}

The existence of a loop $x \to x$ means that the alternative $x \in X$ is `intrinsically bad' in the voter's eyes, and so will never be selected. 
The item eliminates itself as in the case of a person allergic to chocolate: the fact that $x$ contains chocolate deems $x$ ineligible, and this is independent of the structure of $X$ and any possible binary comparison.
On the other hand, a double arrow $x \leftrightarrows y$ between distinct items means that there are scenarios for which a voter considers $x$ strictly better than $y$ ($x \to y$) and different scenarios in which the converse happens ($y \to x$): this yields full indecisiveness, and so the voter will end up choosing none of the two pairwise exclusive items (that is, the ballot associated to this voter displays an empty choice on the menu $\{x,y\}$.)  
 
The difference between Definitions~\ref{DEF:asymmetric_rationalizability} and~\ref{DEF:free_rationalizability} only materializes for DMs whose selection is empty on at least a menu: 

\begin{lemma} \label{LEMMA:representability equals rationalizability for choices} 
\begin{itemize}
	\item[\rm (i)] A choice is asymmetrically rationalizable if and only if it is freely rationalizable.\vs\vs 
	\item[\rm (ii)] If a quasi-choice is asymmetrically rationalizable, then it is freely rationalizable. The converse is false. 
\end{itemize}
\end{lemma}

\noindent \underline{\textsc{Proof}}.
	Asymmetric rationalizability obviously implies free rationalizability.
 	
 	To prove the converse in (i), suppose $c \colon 2^X \to 2^X$ is freely rationalizable by a voter $\to$ over $X$.
  	Since $c(\{x\}) = \{x\}$ for any $x \in X$, the relation $\to$ has no loops.
  	Further, we cannot have $x \leftrightarrows y$ for any distinct $x,y \in X$, since otherwise $c(\{x,y\}) = \es$, a contradiction. 
  	We conclude that the voter $\to$ representing $c$ is asymmetric and loopless, and so $c$ is asymmetrically rationalizable. 
  
  	To complete the proof of (ii), consider the quasi-choice $c$ over $X = \{x,y\}$, defined by $c(A) =A$ if $\vert A \vert =1$, and $c(A)= \es$ otherwise. 
	Then $c$ is freely rationalizable (by the voter $\to$ defined by $x \leftrightarrows y$) but fails to be asymmetrically rationalizable. 
\qed
\bigskip

The asymmetric rationalizability of a choice is characterized by the satisfaction of two classical properties of choice consistency:\vs 
\begin{description}
	\item[Axiom $\alpha\,$:]
	for all $A,B \subseteq X$ and $x \in X$, if $x \in A \subseteq B$ and $x \in c(B)$, then $x \in c(A)$;\vs\vs 
  	\item[Axiom $\gamma\,$:]
    for all $A,B \subseteq X$ and $x \in X$, if $x \in c(A)$ and $x \in c(B)$, then $x \in c(A \cup B)$.\vs
\end{description}

Axiom$\:\alpha$ is due to \cite{Chernoff1954}.
It says that any item selected from a menu is also chosen from any smaller menu containing it: that is why this property is also called \textsl{Standard Contraction Consistency}. 
Its role in the abstract theories of rational individual choice and social choice is central, sometimes in the form of \textsl{Independence of Irrelevant Alternatives} \citep{Arrow1950}. 
The normative appeal of Axiom$\:\alpha$ is rather strong, as discussed in \citet[Remark~1]{EliazOk2006} and \citet[Section~2.2]{Heller2012}. 
In fact, \citet[p.\,407]{Nehring1997} goes even further, calling Axiom$\:\alpha$ ``\textit{the mother of all choice consistency conditions'}'. 
Axiom$\:\gamma$, often referred to as \textsl{Standard Expansion Consistency}, is introduced by \cite{Sen1971}.  
It says that if an item is chosen in two menus, then it is also selected from the larger menu obtained as their union. 

The connection between these properties and rational choice behavior is as follows:

\begin{theorem*}[\citealp{Sen1971}] A choice is asymmetrically rationalizable if and only if both Axiom$\:\alpha$ and Axiom$\:\gamma$ hold.
\end{theorem*}

This characterization extends to quasi-choices, provided that we employ the relaxed notion of rationality described in Definition~\ref{DEF:free_rationalizability}: 

\begin{proposition}[\citealp{AizermanAleskerov1995}, Theorem 2.5] \label{PROP:chrz representability}
  A quasi-choice is freely rationalizable (equivalently, it is a ballot) if and only if Axioms$\;\alpha$ and $\gamma$ hold.
\end{proposition} 

For a proof of Proposition~\ref{PROP:chrz representability}, see \citet[Theorem 2.8]{AleMon2002}.
In connection to Proposition \ref{PROP:chrz representability}, observe that uniqueness of a free rationalization is not guaranteed for quasi-choices: for instance, a null quasi-choice is freely rationalizable by all (and only) the binary relations in which all items are repellent.   


\section{Rationalization by ballots} \label{SECT:multi-rationalizations_by_voters} 

Here we introduce the main notion of the paper, justify it, and characterize it.

\subsection{Preliminaries}

\begin{definition} \rm \label{DEF:s-representability}
	Let $s$ be any number in the interval $[0, 1)$, here called a \textsl{share}. 
	A quasi-choice $c \colon 2^X \to 2^X$ is \textsl{$s$-majoritarian} if there exists a finite family $\mathcal V =\{v_1,\ldots,v_k\}$ of $k$ ballots over $X$, not necessarily pairwise distinct, such that\vs
	\begin{equation} \label{EQ:s-representability}
	 x \in c(A) \quad \Longleftrightarrow \quad \frac{ \big\vert \{i : x \in v_i(A)\} \big\vert }{k} > s\vs	
    \end{equation}	
    for any $A \in 2^X$ and $x \in X$. 
    We say that $c$ is \textsl{rationalizable by ballots} if it is $s$-majoritarian for some share $s \in [0,1)$. 
\end{definition}

In words, $c$ is $s$-majoritarian provided that we can find a finite number of voters such that a candidate is selected from a menu if and only if it is endorsed by a share of voters -- equivalently, a share of the ballots associated with the voters -- strictly larger than $s$. 
It will always be understood that voters/ballots in a given family need not be pairwise distinct. 

The model of rationalization by ballots described in Definition~\ref{DEF:s-representability} aims to build a bridge between choice theory and voting theory.
In fact, a selection process is explained by means of a population $\mathcal V$ of voters, who make a \textit{collective} decision by appealing to a \textit{socially acceptable} paradigm, encoded by a threshold.   

Next, we prove some simple properties of the model given in Definition~\ref{DEF:s-representability}.

\begin{lemma} \label{LEMMA:s-representability}
	Any quasi-choice that is rationalizable by ballots satisfies Axiom$\;\alpha $. 
\end{lemma}

\noindent \underline{\textsc{Proof}}. 
	Suppose $c$ is a quasi-choice over $X$, which is rationalizable by ballots. 
	Thus, there is a family $\mathcal V = \{v_1,\ldots,v_k\}$ of $k$ ballots satisfying~\eqref{EQ:s-representability} for some share $s \in [0,1)$.
	Let $A \subseteq B \subseteq X$ and $x \in A \cap c(B)$.
	By hypothesis, $\frac{| \{i \,: \,x \,\in \, v_i(B) \} |}{k} > s$.
	Since any ballot $v_i$ satisfies Axiom$\;\alpha$ by Proposition~\ref{PROP:chrz representability}, it follows that\vs 
    $$
    \frac{\big\vert \{i :  x \in v_i(A) \} \big\vert}{k} \geqslant \frac{\big\vert \{i  : x \in v_i(B) \} \big\vert}{k} > s,\vs
    $$
    and so $x \in c(A)$ by~\eqref{EQ:s-representability}.
    This proves that $c$ satisfies Axiom$\;\alpha$.
 \qed   
 \medskip
    
The case of a zero share is especially tractable.

%
\begin{lemma} \label{LEMMA:0-representability} A quasi-choice $c$ is $0$-majoritarian if and only if there exists a family $\mathcal V = \{v_1,\ldots,v_k\}$ of ballots such that $c = \bigcup_{i=1}^k v_i$.\footnote{That is, $c(A) = \bigcup_{i =1}^k v_i(A)$ for any menu $A$.}
\end{lemma}

\noindent \underline{\textsc{Proof}}. 
Let $c$ be a quasi-choice over $X$.
By \eqref{EQ:s-representability}, $c$ is $0$-majoritarian if and only if there is a family $\mathcal V =\{v_1,\ldots,v_k\}$ of $k$ ballots over $X$ such that for any $A \in 2^X$,\vs
\begin{align*}
	c(A) = \left\{x \in A : \frac{\big\vert \{i : x \in v_i(A)\} \big\vert}{k} > 0 \right\} = \big\{x \in A : x \in v_i(A) \text{ for some } i\} = \bigcup_{i=1}^k v_i(A).\vs\vs
\end{align*}
This proves the claim. 
 \qed   
 \medskip

Combining Lemma~\ref{LEMMA:0-representability} with Theorems 2.5(a) and 5.4 in~\cite{AizermanAleskerov1995}, we readily derive the following fact:
    	
\begin{corollary} \label{COR:charact-0-representability} Any quasi-choice that satisfies Axiom$\;\alpha $ is {$0$-majoritarian}.
\end{corollary}



\subsection{Interpretation and relation with literature} \label{SECT:liberal and democratic 1} 

To start, let us point out that the standard maximization model is a basic instance of rationalization by ballots. 
Specifically, if a quasi-choice $c$ is freely rationalizable (hence $c$ is a ballot $v_1$ according to Definition~\rm \ref{DEF:free_rationalizability}), then for each $s\in [0, 1)$ the family $\mathcal V = \{ v_1 \} = \{ c \} $ of ballots over $X$ is such that\vs
	\begin{equation} 
	 x \in c(A) \quad \Longleftrightarrow \quad \frac{ \big\vert \{i : x \in v_i(A)\} \big\vert }{1} > s	
    \end{equation}	
    for any $A \in 2^X$ and $x \in X$. 
    
Remarkably, the case in \cite{EliazOk2006} can be expressed in the language of our model, too.
Theorem~2 in the mentioned paper shows that a choice correspondence satisfying the \textsl{Weak Axiom of Revealed Non-Inferiority} \textsl{(WARNI)} can be explained by maximizing (in the sense of Definition~\rm \ref{DEF:free_rationalizability}) the asymmetric part of a reflexive, transitive, and `regular' binary relation.
It follows that all choice correspondences that satisfy WARNI are $s$-majoritarian for each $s\in [0, 1)$. 
In Section~\ref{SECT:liberal and democratic 3} we shall prove that Axiom$\;\alpha$, which is a behavioral consequence of WARNI, is the exact property that justifies the characterization of $s$-majoritarianism for each $s\in [0, 1)$.

Next, consider the model of \cite{KalaiRubinsteinSpiegler2002}, which explains choice \textit{functions} by appealing to a set of binary justifications. 
Specifically, 
%
	a \textsl{rationalization by multiple rationales} \textit{(RMR)} of a choice function $c \colon 2^X \setminus \{\es\} \to X$ is a set $\L$ of linear orders\footnote{A \textsl{linear order} is an asymmetric, transitive and complete binary relation.} (called \textsl{rationales}) over $X$ such that, for all nonempty $A \subseteq X$, the item $c(A)$ is $\succ$-maximal for some $\succ$ in $\L$.     

By definition, any choice function has an RMR: assign to each menu a linear order having the selected item as its top element. 
As a consequence of the non-testability of the RMR model,  the authors study the `degree of rationality' of choice functions, intended as the minimum number of rationales that are needed to justify selection on each menu.
Then, the larger this degree of rationality is, the less rational a choice behavior is labeled. 
(Note that this degree is $1$ if and only if the choice is rationalizable.) 
The authors show that a choice function on $n$ elements can be explained by $n-1$ rationales; however, the proportion of choices that require the maximum number of rationales tends to one as the number of items tends to infinity. 
Thus, distinguishing choice functions by their degree of rationality fails to be very effective, especially for large grand sets.  
%

\begin{remark} \rm \label{REM:KRS} 
One could also define rationalization by multiple rationales of a choice \textsl{correspondence} $c$ as a set $\W$ of weak orders\footnote{A \textsl{weak order} is an asymmetric and negatively transitive (hence transitive) binary relation.}
over $X$ such that, for all nonempty $A \subseteq X$, $c(A)$ is the set of $\succ$-maximal elements for some $\succ$ in $\W$.		
Then any choice correspondence has a rationalization by multiple rationales, too.
An analysis of the minimum number of rationales that are needed to guarantee rationalization appears to be interesting, but is beyond the scope of this paper.
\end{remark}

Apart from some drawbacks already put across by the authors (see, e.g., \citealp{KalaiRubinsteinSpiegler2002}, p.\,2847), we detect an additional setback of this model: the same RMR can simultaneously explain rather different choice functions.

\begin{example} \rm \label{EX:choices with common multi-rationalization}
Let $c_1,c_2 \colon 2^X \setminus \{\es\} \to X$ be the choice functions over $X=\{x, y, z\}$ respectively defined by\vs\vs
$$
(c_1)\;\; \underline{x}{y}\,,\;{x}\underline{z}\,,\;y\underline{z}\,,\;\underline{x}yz \qquad\text{and}\qquad   
(c_2)\;\; \underline{x}{y}\,,\;\underline{x}{z}\,,\;y\underline{z}\,,\;xy\underline{z}\,.\vs\vs
$$   
Both $c_1$ and $c_2$ are non-rationalizable, and can be explained by the same set of two rationales.       
Specifically, let $\succ_1$ and $\succ_2$ be the linear orders over $X$ such that $x \succ_1 y \succ_1 z$ and $z \succ_2 x \succ_2 y$.
We claim that $\L = \{\succ_1,\succ_2\}$ is an RMR for both $c_1$ and $c_2$.
Indeed, for $c_1$, the linear order $\succ_1$ rationalizes the menus $X$ and $\{x,y\}$, whereas $\succ_2$ rationalizes $\{x,z\}$ and $\{y,z\}$.
On the other hand, for $c_2$, the linear order $\succ_1$ rationalizes the menus $\{x,y\}$ and $\{x,z\}$, whereas $\succ_2$ rationalizes $\{y,z\}$ and $X$.
\end{example}

The fact that the same RMR may be employed to explain very different selections is a shortcoming of the described model.
However, this can be easily fixed, rewriting the definition of an RMR according to a `functional approach'.  
Specifically, let a \textsl{rationalizer for $c$} be a pair $\langle \L, f\rangle$ such that $\L$ is a family of linear orders over $X$, and $f \colon 2^X \setminus \{\es\} \to \L$ is a map such that $c(A) = \max(A,f(A))$ for all $A \in 2^X \setminus \{\es\}$.   
The function $f$ assigns a rationale to each menu, and so the two choices in Example~\ref{EX:choices with common multi-rationalization} now do have distinct rationalizer.  
The functional approach suggested for the RMR approach is, in fact, the one that we shall employ -- \textit{mutatis mutandis} -- in our model. 

Alternative interpretations of a rationalization by ballots are related to other bounded rationality approaches.
Multi-self decision making -- see, e.g., \cite{May1954} and \cite{ManziniMariotti2007} -- provides interpersonal or intrapersonal frameworks for aggregation of preferences.
The essential idea of this paradigm is that in order to derive choices over alternatives, a DM resorts to a family $\mathcal V$ of fictitious selves (also called `motivations' or `priorities'). 
In this respect, here we employ the suggestive terms `ballots' and `voters' instead, and do not impose an interpersonal interpretation.
We also assume that each ballot $v \in \mathcal V$ can be generally indecisive, including indecisiveness for singletons and doubletons. 
Even more important, we consider the general case of quasi-choice correspondences instead of the very restricted scenario of choice functions. 
From this point of view, a quasi-choice correspondence is rationalizable by ballots whenever there is a family $\mathcal V$ of (possibly fictitious or unobservable) voters with the property that the items selected from each menu are exactly those endorsed by more than a fixed share of voters.  

Another interpretation of our model is related to \textsl{multiple criteria decision making}.
In such a scenario, alternatives are characterized by attributes, and each attribute is explained by a rationale. 
When the DM only attends to selected characteristics, she makes a choice by the attributes that stem from the maximization of the corresponding rationales.  
Choices may be justified by these `selections by attributes' as follows: a DM decides to accept an alternative from a menu when it is choosable for a fixed share of attributes. 
Definition~\ref{DEF:s-representability} considers choices that can be explained `as if' the DM were assigning attributes to the set of options, and in that frame of mind her choices are justified by at least a certain number of attributes. 
%
%


\subsection{Characterization} \label{SECT:liberal and democratic 3}

Since a share $s$ in Definition~\ref{DEF:s-representability} is, by its own nature, a proportion of a finite society, we may assume that $s$ is a rational number.  
In such a setting we have, as announced:

\begin{theorem} \label{THM:main1}
			The following statements are equivalent for a quasi-choice $c\colon 2^X \to 2^X$:\vs\vs 
\begin{itemize}
	\item[\rm(i)]  $c$ satisfies Axiom$\;\alpha$;\vs\vs
	\item[\rm(ii)] $c$ is $s$-majoritarian for some rational share $s \in [0,1)$;\vs\vs
	\item[\rm(iii)] $c$ is $s$-majoritarian for each rational share $s \in [0,1)$.
\end{itemize}
\end{theorem}

Theorem~\ref{THM:main1} suffices for practical purposes. 
However, we wish to point out that it also holds true for irrational shares: 

\begin{corollary} \label{COR:full_generality}
			The following statements are equivalent for a quasi-choice $c \colon 2^X \to 2^X$:\vs\vs 
\begin{itemize}
	\item[\rm(i)]  $c$ satisfies Axiom$\;\alpha$;\vs\vs 
	\item[\rm(ii)] $c$ is $s$-majoritarian for some share $s \in [0,1)$;\vs\vs
	\item[\rm(iii)] $c$ is $s$-majoritarian for each share $s \in [0,1)$. 
\end{itemize}
\end{corollary}

Theorem~\ref{THM:main1} and Corollary~\ref{COR:full_generality} are proved in the Appendix.
They assert that all $s$-majoritarian justifications are `equally sound', because they are witnessed by the same behavioral property: either all of them are valid explanations of a given choice behavior, or none is.

However, this equivalence vanishes as soon as we consider the simplest feature of a justification by ballots, namely the number of voters that is required in the group.
Section~\ref{SECT:numbers} investigates this issue by focussing on two benchmark specifications of $s$-majoritarianism, where the share is either one half (`democracy') or zero (`liberalism').  


\section{Group size matters: distinguishing majoritarian justifications} \label{SECT:numbers}

Rationalization by ballots is a parametric choice model, with the share $s \in (0,1]$ being the parameter. 
For a fixed value of $s$, the complexity of any $s$-majoritarian justification is measured by the minimum number of voters needed for it.
In this perspective, for any observed (possibly non-decisive) choice behavior satisfying Standard Contracting Consistency, it may be interesting to determine the share $s$ which yields the $s$-majoritarian justification with the least minimum number of voters.
This would entail the `least complex' explanation within the majoritarian paradigm, also providing a novel measure of the degree of rationality of quasi-choices.
%
%

Apparently, the general version of this problem appears to be rather challenging; thus, we leave it open.
However, we do have some partial results, which allow one to gauge its difficulty.
Specifically, in this section we shall investigate the minimum size of populations witnessing two benchmark types  of majoritarian justifications: democratic and liberal. 

\begin{definition} \rm 
Let $c \colon 2^X \to 2^X$ be a quasi-choice over $X$ satisfying Axiom$\:\alpha$.
A \textsl{democratic representation of $c$ of size $k$} is any family $\mathcal V =\{v_1,\ldots,v_k\}$ of $k$ ballots over $X$ such that for any $A \in 2^X$ and $x \in X$,\vs\vs
	\begin{equation} \label{EQ:democratic choice} 
		 x \in c(A) \quad \Longleftrightarrow \quad \big\vert \{i : x \in v_i(A)\} \big\vert > \frac{k}{2}.\vs\vs
    \end{equation}
Similarly, a \textsl{liberal representation of $c$ of size $k$} is any family $\mathcal V =\{v_1,\ldots,v_k\}$ of $k$ ballots over $X$ such that for any $A \in 2^X$ and $x \in X$,\vs\vs
	\begin{equation} \label{EQ:liberal choice} 
		 x \in c(A) \quad \Longleftrightarrow \quad x \in v(A) \text{ for some } v_i \in \mathcal V.
    \end{equation}
\end{definition}

As a first evidence of the difficulty of the problem posed above, in Section~\ref{SUBSECT:relative_bounds_for_numbers} we shall prove a rather counterintuitive result, in relation to the comparison of the complexity of democratic and liberal explanations.

\subsection{Democratic vs liberal representations} \label{SUBSECT:relative_bounds_for_numbers}

The proof of Theorem~\ref{THM:main1} suggests that the minimum number of ballots/voters needed to achieve a democratic representation is typically different from that needed to obtain a liberal one. 
This observation yields a natural proxy that differentiates these two ways of justifying choices by ballots:

\begin{definition} \rm
  For any quasi-choice $c \colon 2^X \to 2^X$, we define the \textsl{democratic number} $\dem(c)$ as the size of a smallest democratic representation of $c$, if there are any, and infinite otherwise, namely \vs\vs
   $$
     \dem(c) := \left\{
        \begin{array}{llll}
          \hbox{size of smallest democratic representation of $c$}  
          & \hbox{ if } c \hbox{ satisfies Axiom$\:\alpha$,} \\
          \infty & \hbox{ otherwise.}
        \end{array}
  \right.\vs
  $$
Similarly, the \textsl{liberal number} $\lib(c)$ is defined as the size of a smallest liberal representation of $c$, if there are any, and infinite otherwise, namely\vs\vs
  $$
     \lib(c) := \left\{
        \begin{array}{llll}
          \hbox{size of smallest liberal representation of $c$} 
          & \hbox{ if } c \hbox{ satisfies Axiom$\:\alpha$,}  \\
          \infty & \hbox{ otherwise.}  
        \end{array}
  \right. 
  $$
\end{definition}

Thus, we have:\vs\vs
\[
\dem(c) < \infty \quad \Longleftrightarrow \quad \hbox{$c$ satisfies Axiom$\:\alpha$} \quad \Longleftrightarrow \quad \lib(c) < \infty.\vs
\]
\medskip 

In a situation of full rationality, the two paradigms are indistinguishable: 

\begin{lemma} \label{LEMMA:lib and dem for rationalizable}
A quasi-choice $c$ is freely rationalizable if and only if $\dem(c) =1$ if and only if $\lib(c) =1$.
\end{lemma}

Lemma~\ref{LEMMA:lib and dem for rationalizable} follows from 
the fact that a democratic representation of $c$ by a single (arbitrary) binary relation is also a liberal representation of $c$. 

However, for non-rationalizable quasi-choices, the liberal and the democratic paradigms may display a different behavior concerning the minimal size of a representation, even on small sets of alternatives.
The next two examples provide some simple instances of this kind, moreover in the case of decisive choices.

\begin{example} \rm  \label{EX:lib=2 dem=3} 
 Let $c$ be the choice over $X=\{x, y, z\}$ defined by $\underline{x}\underline{y}\,,\;\underline{x}z\,,\;\underline{y}z\,,\;\underline{x}yz\,.$ 
  Note that $c$ satisfies Axiom$\;\alpha$ but not Axiom$\;\gamma$, and so fails to be rationalizable.
  In the Appendix we prove that $\lib(c) = 2$ and $\dem(c) = 3$.
  \end{example}

\begin{example} \label{EX:dem less than lib} \rm 
	Let $c$ be the choice over $X = \{0,1,2,3,4,5\}$ defined by\vs\vs 
	$$
	c(A) \; = \; \left\{
	\begin{array}{lll}
	  A \setminus \{0\} & \hbox{ if } 0 \in A \hbox{ and } \vert A \vert \geq 4 \\
	  A & \hbox { otherwise.} 
	\end{array} 
	\right.\vs\vs 
	$$
	Again, $c$ satisfies Axiom$\;\alpha$ but not Axiom$\;\gamma$, which implies $2 \leqslant \dem(c) < \infty$ and $2 \leqslant \lib(c) < \infty$. 
    In the Appendix we prove that $\dem(c) \leq 5$ and $\lib(c) \geq 10$.
    Note that the liberal and democratic paradigms are \textit{very} far apart in this case.
    This can be interpreted as saying that a situation in which a government only disregards a single item when the latter appears in a large menu (otherwise being fully neutral) is better suited for a democratic model rather than a liberal one.  
\end{example}

On the other hand, the democratic and liberal numbers are equal for some non-rationalizable choices:

\begin{example} \label{EX:dem = lib} \rm  
	Let $c$ be the choice over $X=\{x, y, z, w\}$ defined by\vs\vs
  $$
  \underline{x}\underline{y}\,,\;\underline{x}\underline{z}\,,\;\underline{x}w\,,\;\underline{y}\underline{z}\,,\;\underline{y}w\,,\;\underline{z}\underline{w}\,,\quad\underline{x}\underline{y}z\,,\;\underline{x}\underline{y}w\,,\; \underline{x}zw\,,\;\underline{y}zw\,,\quad\underline{x}yzw\,.\vs\vs
  $$	
  Again, Axiom~$\alpha$ holds for $c$, but Axiom~$\gamma$ does not. 
  It follows that $2 \leqslant \dem(c)< \infty$ and $2 \leqslant \lib(c) < \infty$. 
  In the Appendix we prove that $\dem(c) = \lib(c) = 3$.
\end{example}

We hinted above that some bounded rationality models define the complexity of an observed choice behavior by representing it in terms of a parametric model, and interpreting the minimum number of parameters needed for a representation as a measure of this complexity: \citet{KalaiRubinsteinSpiegler2002} and \citet{AmbrusRozen2014} provide two notable examples of this kind.
However, this type of approach raises a challenging philosophical question: \textit{How convincing is the underlying parametric representation as an actual model of decision-making?}

In this respect, our parametric model is not designed with the exclusive purpose of measuring the complexity of choice behavior by  appealing to the minimum number of voters needed for  an $s$-majority.
Regrettably, Examples~\ref{EX:lib=2 dem=3}-\ref{EX:dem = lib} yield a possibly counterintuitive conclusion, because  even a comparison of the complexity of the $0.5$-majority and  $0$-majority explanations fails to hold true.
In view of this setback, our goal is first to produce concrete evidences that tell the liberal paradigm apart from the democratic one, and only then measuring how distant these two specifications of the $s$-majoritarian model are from each other. 

All in all, we can distinguish three types of quasi-choices for which Axiom~$\alpha$ holds: (i) those with a liberal number  smaller than the democratic number (see Example~\ref{EX:lib=2 dem=3}); (ii) those for which the reverse happens (see Example~\ref{EX:dem less than lib}); (iii) those such that the two numbers are equal (see Example~\ref{EX:dem = lib}).  
A systematic analysis of the democratic and liberal numbers of a quasi-choice satisfying Axiom~$\alpha$ appears to be non-trivial.
We leave open the problem of determining a \textit{testable} characterization of the classes (i)--(iii). 
Finding an $s$-majoritarian justification with a smallest minimum number of voters, for each quasi-choice satisfying Axiom~$\alpha$, is another open problem that we mentioned above in this section.

The following result provides some `relative' upper bounds, that is, bounds for the democratic (resp.\ liberal) number in terms of the liberal (resp.\ democratic) number:

\begin{lemma} \label{LEMMA:respective upper bounds}
	For any quasi-choice $c$, we have:\vs 
	\begin{itemize}
		\item[\rm (i)] $\dem(c) \leqslant 2\, \lib(c)$;\vs 
		\item[\rm (ii)] $\lib(c) \leqslant 2^{\,\dem(c) -1}$.
	\end{itemize}
\end{lemma}

Lemma~\ref{LEMMA:respective upper bounds} partially answers the following question: \textit{If we have a liberal (resp.\ democratic) representation of a possible indecisive choice behavior, how large should the population of voters be in order to obtain a democratic (resp.\ liberal) justification for the same behavior?}
In this respect, the proof of Lemma~\ref{LEMMA:respective upper bounds} is instructive, because it explicitly shows how to canonically construct a liberal (resp.\ democratic) representation from a democratic (resp.\ liberal) one. 


\subsection{Asymptotic behavior of liberalism} \label{SUBSECT:asymptotic_behavior}

We conclude this section by establishing an upper bound for the liberal number of any quasi-choice that satisfies Axiom$\:\alpha$ (in relation to the size of the grand set). 
Since this bound is tight, we shall also get the asymptotic behavior of $\lib(c)$.

To that end, we need a combinatorial result (Lemma~\ref{cnk-lemma}), in which we abstract the main ingredients in the proof of Theorem~\ref{COR:full_generality}.
In it, we construct an infinite collection $\{c_{n,k} : n \in \mathbb{N} ,\,1 \leqslant k \leqslant n+1 \}$ of choices over a grand set $\{ 0, 1, \ldots, n \}$ of size $n+1$ having the following features:\vs\vs 
\begin{itemize}
	\item all of them satisfy Axiom$\:\alpha$ (hence they are $s$-majoritarian for any $s \in [0,1)$);\vs\vs 
	\item they have a fixed (but quite `large') liberal number;\vs\vs 
	\item they have a democratic number that is bounded above (by a `small' integer).\vs\vs 
\end{itemize}
These choices are very close to being neutral, because all elements, with the possible exception of the item $0$, are selected in all menus.

\begin{lemma} \label{cnk-lemma}
For each $1 \leqslant k \leqslant n+1$, let $c_{n,k}$ be a choice over the nonempty grand set $X_{n} = \{0,1,\ldots,n\}$ satisfying the following conditions for all nonempty $A \in 2^X$:\vs\vs
\begin{enumerate}[label={\rm(\alph*)}]
	\item\label{cnk-lemmaa} $A \setminus c_{n,k}(A) \subseteq \{0\}$;\vs\vs
	\item\label{cnk-lemmab} $A \setminus c_{n,k}(A) = \{0\}$ if and only if $0 \in A \hbox{ and } |A| > k$.\vs\vs
\end{enumerate}
Then $c_{n,k}$ satisfies Axiom$\:\alpha$. 
Furthermore,\vs\vs  
\begin{equation} \label{EQ:cnk-lemma}
	\lib(c_{n,k}) = {n \choose k-1} \qquad \text{and} \qquad
	\begin{cases}
	\dem(c_{n,k}) \leqslant 2k-1 & \text{if~ } \frac{n}{2} < k \leqslant n+1\\
	\dem(c_{n,k}) \leqslant 2(n-k) & \text{if~ } 1 \leqslant k \leqslant \frac{n}{2} .
	\end{cases}
\end{equation}
\end{lemma}

From Lemma~\ref{cnk-lemma}, we obtain what we were after: 

\begin{theorem} \label{THM:upper bound lib}
For any quasi-choice $c$ over a grand set of size $n \geqslant 2$ satisfying Axiom~$\alpha$,\vs\vs
\begin{equation}\label{EQ:upper bound lib}
\lib(c) \,\leqslant\, {n -1 \choose \lfloor \frac{n-1}{2} \rfloor}.\vs\vs
\end{equation}
Moreover, the upper bound~\eqref{EQ:upper bound lib} is tight.  
\end{theorem}
\medskip

Finally, we determine the asymptotic behavior of $\lib(c)$.
Since Stirling's approximation entails\vs
$$
{n -1 \choose \lfloor \frac{n-1}{2} \rfloor} \,=\, \mathcal{O} \left(\frac{2^{n}}{\sqrt{n}} \right),
$$
the tightness of the upper bound in Theorem~\ref{THM:upper bound lib} yields the following result:

\begin{corollary} \label{COR:limit behavior of lib}
For any quasi-choice $c$ over a grand set of size $n \geqslant 2$ satisfying Axiom$\;\alpha$,\vs\vs 
$$
\lib(c) \,=\, \mathcal{O}\left(\frac{2^{n}}{\sqrt{n}}\right).
$$
\end{corollary}


\section{Final remarks} \label{SECT:conclusions}

Several models show that effective theories of individual choice can be founded on tenets that relax the classical Weak Axiom of Revealed Preference, possibly in combination with some other common postulates.
Our approach to associate observed choice behavior with maximizing attitudes is `radical', because it addresses a root concern: we identify a liberal maximization process that can be tested solely by Standard Contraction Consistency. 
Moreover, in this liberal approach, we have explored the justifications that require the minimal number of voters.
A less radical justification of the same class of choices places them in a majoritarian scenario with respect to an arbitrarily large share.
Here our numerical conclusions are less complete.
Our procedure is in continuation of the strand of literature on rationalizability of choice with multi-self models that started with \cite{May1954}.

 Alternatively, when choices are single-valued, \cite{ManziniMariotti2007} explain some empirically important `irrational' patterns of choice by a \textsl{rational shortlist method (RSM)}.
`Irrational' means that basic rationality tenets, such as Chernoff's Axiom~$\alpha$ and Sen's Axiom~$\gamma$, are contradicted.
In a RSM, a first rationale (an asymmetric binary relation) gives a selection of alternatives (a shortlist); then a second rationale (another asymmetric binary relation) determines the unique selection.
Notice that there are choice functions that are neither RSM nor $s$-majoritarian, because they violate Axiom$\:\alpha$.  
A simple example of this kind is the `default route' example in ~\citet[Section I.B]{ManziniMariotti2007}.
Indeed, for $X=\{A,B,C\}$, the (single-valued) choice function defined by $A\underline{B}$, $\underline{A}C$, $\underline{B}C$, and $\underline{A}BC$ is not an RSM; moreover, since Axiom$\:\alpha$ fails to hold, this choice fails to have an $s$-majoritarian representation.
			
Rationalization by multiple rationales \citep{KalaiRubinsteinSpiegler2002} does not use rationales in a sequential order.
Our model  resorts to a multiplicity of rationales but it does not use them sequentially either.
As in an RMR approach, one rationale justifies the choice of an item in the liberal (i.e., $0$-majoritarian) model.
Unlike that model, however,  neither of the rationales is irrelevant, and we allow for two framework effects, because both the menu and the item may affect the mental argument -- the particular \textit{self} -- that acts for justification. 
Furthermore, the introduction of a new voter (a bounding rationale) modifies the choice.
Choices are not necessarily unique, and we need not be explicit about the interpretation of multiple choices.

\cite{EliazOk2006} show how a choice theory that allows `rational agents' to remain indecisive at times can be founded on WARNI.
Since Axiom$\:\alpha$ is a behavioral consequence of WARNI, our model places their `rational agents' in a (specific) multi-selves context \citep[Footnote 29]{EliazOk2006}. 

One issue remains unexplored: computational complexity.
With respect to the RMR model, \cite{ApesteguiaBallester2010} and afterwards \cite{Demuynck2011} show that the problems of rationalizing choices by a minimum number -- or by a fixed number -- of rationales are $\textbf{NP}$-complete.
The class of $\textbf{NP}$-complete problems consists of the most difficult $\textbf{NP}$ problems.
In turn, any $\textbf{NP}$ problem is deemed `easy to verify' (i.e., there are algorithms that check in polynomial time whether a purported solution is indeed a solution), although it might be `difficult' to find a solution (e.g., because solving it might take exponential time).
Also the problem posed by sequential choice by $K$ rationales \citep{ManziniMariotti2007} is $\textbf{NP}$-complete when $K\geq 3$ \citep[Theorem 3]{Demuynck2011}.
In the opposite direction, the latter result proves that the problem of whether a choice correspondence is explained by the status-quo bias model \citep{MasatliogluOk2005} is decidable in polynomial time (i.e., it is in the class $\textbf{P}$ of decision problems).
By exploiting the bounds in Theorem~\ref{THM:upper bound lib} and  Lemma~\ref{LEMMA:respective upper bounds}(i), it is not difficult to show that checking whether $\lib(c) \,\leq \, k$ is true, or $\dem(c) \,\leq \, k$, is an $\textbf{NP}$ problem.
 This means that the answer can be verified in polynomial time.
 However, we do not know whether the problems $\lib(c) \,\leq \, k$ and $\dem(c) \,\leq \, k$ are also $\textbf{NP}$-hard (hence $\textbf{NP}$-complete) or not.
 The bound established by Corollary~\ref{COR:limit behavior of lib} does not help for this purpose.
	

\section*{Appendix: Main Proofs}

\noindent \underline{\textsc{Proof of Theorem~\ref{THM:main1}}}. 
Since (iii)$\;\Longrightarrow\;$(ii) is obvious, and (ii)$\;\Longrightarrow\;$(i)  is Lemma~\ref{LEMMA:s-representability} (note that $s$ is not required to be a rational number), we only show (i)$\;\Longrightarrow\;$(iii).
 \medskip
      
\noindent \textbf{(i)$\;\Longrightarrow\;$(iii):} 
Suppose $c$ satisfies Axiom~$\alpha$, and let $s \in \QQ \cap [0,1)$. 
We prove the claim in three steps:\vs
\begin{enumerate}  
	\item $c$ is $0$-majoritarian;\vs
	\item $c$ is $t$-majoritarian for some $t\in (s, 1)\cap \QQ$;\vs  
	\item  $c$ is $s$-majoritarian.\vs
\end{enumerate}

\begin{description}
	\item[Step 1:]  
This is Corollary~\ref{COR:charact-0-representability}, which owes to \cite{AizermanAleskerov1995}, Theorems~2.5(a) and 5.4. 
Let $\mathcal V = \{v_1,\ldots, v_n\}$ be a finite family of ballots witnessing the fact that $c$ is $0$-majoritarian.\footnote{Recall that we allow ballots in $\mathcal V$ to be equal.} 

\item[Step 2:] We build on the family $\mathcal V$ of ballots given at Step~1.
Since the sequence $\left\{\frac{k^2+k-1}{k^2+nk}\right\}_{k =1}^\infty$ converges to 1, and $\frac{k^2+k-1}{k^2+n k}<1$ for any $k \geqslant 1$, there exists $m\in \NN$ such that $s < \frac{m^2+m-1}{m^2+nm} < 1$.
Set $t:=\frac{m^2+m-1}{m^2+nm}\in \QQ \in (0,1)$.
We claim that $c$ is $t$-majoritarian. 

To prove the claim, we construct a finite family $\mathcal W$ of ballots witnessing that $c$ is $t$-majoritarian. 
Let $\mathcal V'$ be a family of `neutral' ballots having size $|\mathcal V'| = m^2$, that is, for each $v' \in \mathcal V'$, the equality $v'(A) = A$ holds for all menus $A \subseteq X$.\footnote{Since $\mathcal V'$ is a family (and not a set, because elements can be repeated), the size $\vert \mathcal V' \vert$ of $\mathcal V'$ is the cardinality of the index set.} 
Furthermore, let $\mathcal V''$ be the family of ballots obtained by replicating $m$ times each ballot in the original family $\mathcal V$.
Set $\mathcal W := \mathcal V' \cup \mathcal V''$, hence $\mathcal W$ is made of $m^2 + mn$ ballots. 
To check that $\mathcal W$ is a $t$-majoritarian representation of $c$, let $a \in A \subseteq X$.   
The forward implication in \eqref{EQ:s-representability} holds, because if $a \in c(A)$, then\vs
\begin{align*} 
	 \frac{ \big\vert \{w  \in \mathcal W : a \in w(A)\} \big\vert }{\vert W \vert} \;& = \; 
	 \frac{\big\vert \{v' \in \mathcal V' : a \in v'(A)\} \big\vert + m \cdot \big\vert \{v \in \mathcal V : a \in v(A)\} \big\vert}{\vert \mathcal W \vert} \\
	 & \geqslant \; \frac{m^2 + m}{m^2 + mn} \; > \; t.
\end{align*} 
For the converse, suppose $\frac{\vert \{w \,\in\, \mathcal W \,:\, a \,\in\, w(A)\} \vert}{\vert \mathcal W \vert} > t$.
Then there is $v \in \mathcal V$ such that $a \in v(A)$, since otherwise $\frac{\vert \{w \,\in\, \mathcal W \,:\, a \,\in\, w(A)\} \vert}{\vert \mathcal W \vert} = \frac{m^2}{m^2 +mn} < t$. 
Since $\mathcal V$ witnesses that $c$ is $0$-majoritarian, we obtain $a \in c(A)$. 
This proves that \eqref{EQ:s-representability} holds. 

\item[Step 3:] We suitably modify the family $\mathcal W$ defined at Step 2. 
For brevity, denote by $p := m^2 + mn$ the size of the family $\mathcal W$. 
First we argue that we can assume that the number $p' := p (\frac{t}{s}-1)$ is a positive integer. 
Indeed, $\frac{t}{s}-1$ is a positive rational number, hence a sufficiently large replication of $\mathcal W$ satisfies $p' \in \NN$.
(Observe that all replications of $\mathcal W$ are obviously $t$-majoritarian justifications of $c$.)
Now consider a family $\mathcal W'$ of $p'$ `hypercritical' ballots (that is, ballots such that the selection from each menu is always empty).    
Finally, set $\mathcal Z := \mathcal W \cup \mathcal W'$.
Note that $\vert \mathcal Z \vert = p + p' = p \cdot \frac{t}{s}$. 

We check that $\mathcal Z$ is an $s$-majoritarian representation of $c$.
Indeed, for each $a\in X$, we have\vs\vs
\begin{equation*} 
\begin{aligned}
	 \frac{\big\vert \{z \in \mathcal Z : a \in z(A)\} \big\vert }{\big\vert \mathcal Z \big\vert} \; & = 
	 \;  \frac{\big\vert \{w \in \mathcal W : a \in w(A)\} \big\vert}{p \cdot \tfrac{t}{s}} & = 
	 \; \frac{\big\vert \{w \in \mathcal W : a \in w(A)\} \big\vert}{\big\vert \mathcal W \big\vert} \cdot \frac{s}{t}.
\end{aligned}\vs\vs
\end{equation*} 
It follows that\vs\vs
\begin{equation*} 
\begin{aligned}
	 \frac{ \big\vert \{z \in \mathcal Z : a \in z(A)\} \big\vert}{\big\vert \mathcal Z \big\vert} >s 
	 \quad & \Longleftrightarrow \quad  
	 \frac{\big\vert \{w \in \mathcal W : a \in w(A)\} \big\vert}{\big\vert \mathcal W \big\vert} > t.
\end{aligned}\vs\vs
\end{equation*} 
\end{description}
This completes the proof of Theorem~\ref{THM:main1}. 
\qed
\bigskip


%
\noindent \underline{\textsc{Proof of Corollary~\ref{COR:full_generality}}}.
We have already observed that the proof of (ii)$\;\Longrightarrow\;$(i) in Theorem~\ref{THM:main1} goes through without requiring the share $s$ be rational.  
Thus, it suffices to show that (i)$\;\Longrightarrow\;$(iii) holds. 
\smallskip

To that end, suppose $c$ satisfies Axiom$\:\alpha$, and let $s \in (0,1)$.  
Select $t \in \QQ$ such that $s<t<1$.
By Theorem \ref{THM:main1}, $c$ is $t$-majoritarian; let $\mathcal V$ be a family of $p$ ballots over $X$ witnessing that $c$ is $t$-majoritarian.
By replicating the society as many times as needed, we can assume without loss of generality that $p \geqslant 2$, $p  t  \in \NN$, and $t<\frac{p-1}{p}$. 
 
Consider the sequence $\left\{t -\frac{tk}{p+k}\right\}_{k=0}^{\infty}$, which converges to $0$ as $k$ diverges. 
We claim that the inequality
\begin{equation} \label{EQ:help_corollary}
t-\frac{t(k-1)}{p+(k-1)} \;<\; t-\frac{tk-1}{p+k}\vs
\end{equation}
holds for all $k \in \NN$. 
Since $t<\frac{p-1}{p}$, the claim holds for $k=0$.
Routine computations show that \eqref{EQ:help_corollary} is equivalent to $p(t-1)+1-k<0$, which is true for $k\geqslant 1$. 
This proves \eqref{EQ:help_corollary}. 
Let $q$ be the minimum positive integer such that $t-\frac{tq}{p+q} \leqslant s$.
Since\vs\vs
$$
t-\frac{tq}{p+q} \;\leqslant\; s \;<\; t -\frac{t(q-1)}{p+q-1} \;<\; t -\frac{tq-1}{p+q},\vs\vs
$$
there is $\delta \in [0,1)$ such that $s = t-\frac{tq-\delta }{p+q} = \frac{\left(t + \frac{\delta }{p}\right)p}{p+q}$.
To complete the proof, we show that $c$ is $s$-majoritarian in two steps.
\begin{description}
	\item[Step 1:] \textit{$c$ is $\left(t + \frac{\delta}{p}\right)$-majoritarian.} 
	By assumption, for any $A \in 2^X$, we have\vs\vs
		 $$
		 a \in c(A) \quad \Longleftrightarrow \quad  \big\vert \{v \in \mathcal V : a \in v(A)\} \big\vert  > t  |\mathcal V|.\vs\vs
		 $$
	Furthermore,\vs\vs 
	$$
	\big\vert \{v \in \mathcal V : a \in v(A)\} \big\vert \;\geqslant\; t  |\mathcal V| + 1 \;>\; t | \mathcal V| + \delta \;\geqslant\; t |\mathcal V|,\vs\vs
	$$ 
	since $\big\vert \{v \in \mathcal V : a \in v(A)\} \big\vert$ and $t |\mathcal V|$ are integers.
	Thus the claim follows from\vs\vs
	$$
	a \in c(A) \quad \Longleftrightarrow \quad \big\vert \{v \in \mathcal V : a \in v(A)\} \big\vert  > t |\mathcal V| + \delta = t p + \delta.	
	$$
	\item[Step 2:] \textit{$c$ is $s$-majoritarian.} 
	Let $\mathcal V'$ be a family of $q$ hypercritical ballots (empty quasi-choices) over $X$. 
	Set $\W := \mathcal V \cup \mathcal V'$, hence $|\W| = p + q$. 
	For each $A \in 2^X$, we have\vs\vs  
	\begin{align*}
		a \in c(A) \quad 
		 & \Longleftrightarrow \quad 
		 \frac{\big\vert \{v \in \mathcal V : a \in v(A)\} \big\vert}{p} > t + \frac{\delta}{p}\\	
		 & \Longleftrightarrow \quad
		 \frac{\big\vert \{w \in \W : a \in w(A)\} \big\vert }{p} > t + \frac{\delta }{p}\\
		 & \Longleftrightarrow \quad 		 
		 \frac{\big\vert \{w \in \W : a \in w(A)\} \big\vert }{|\W|} > \frac{(t + \frac{\delta }{p})p}{p+q} = s.\vs\vs
	\end{align*}
	Thus the family $\W$ provides a $s$-majoritarian justification of $c$, as claimed. \qed
\end{description}
\bigskip
%
%

%

\noindent \underline{\textsc{Proof of Example~\ref{EX:lib=2 dem=3}}}.
  We know that the choice $c$ fails to be asymmetrically rationalizable. 
  By Lemma~\ref{LEMMA:representability equals rationalizability for choices}(i), $c$ is not freely rationalizable either.
  Lemma~\ref{LEMMA:lib and dem for rationalizable} and Theorem~\ref{COR:full_generality} yield $2 \leqslant \dem(c)< \infty$ and $2 \leqslant \lib(c) < \infty$. 
   We show that $\lib(c) =2$ and $\dem(c) =3$ by explicitly exhibiting the binary relations (voters) that rationalize $c$. 
   \begin{description}
   	\item[\rm $\lib(c)=2\,$:] 
   		Let $\to_1$ and $\to_2$ be the two binary relations over $X$ defined by\vs\vs
   		\begin{itemize}
   			\item[(1)] $x \to_1 y$, $x \to_1 z$, and $y \to_1 z$ (that is, $\to_1$ a strict linear order over $X$),\vs\vs
   			\item[(2)] $x \to_2 z$ and $y \leftrightarrows_2 z$.\vs\vs
   		\end{itemize}   
   		One can readily check that the set $\{\to_1,\to_2\}$ is a liberal representation of $c$. 
   \item[\rm $\dem(c)=3\,$:]
 		Let $\to_3$ be the empty relation, which implies that the associated ballot $c_3$ is the neutral choice (i.e., $c_3(A) = A$ for any menu $A \subseteq X$). 
 		One can readily check that $\{\to_1,\to_2,\to_3\}$ is a democratic representation of $c$.
 		This proves $\dem(c) \leqslant 3$.  
 		To complete the proof, we show that no democratic representation of $c$ may have size two. 
 		Toward a contradiction, suppose $\{\to^\prime_1,\to_2^\prime\}$ is a democratic representation of $c$. 
 		Since $c(\{x,y\}) = \{x,y\}$, the democratic paradigm implies that $x \not\rightarrow_i^\prime x,y$, and $y \not\rightarrow^\prime_i x,y$ for all $i \in \{1,2\}$. 
 		Moreover, from $c(\{x,z\}) = \{x\}$ and $c(\{y,z\}) = \{y\}$ we deduce that $z \not\rightarrow_i^\prime x,y$ for all $i \in \{1,2\}$. 
 		We conclude $x,y \in \max(X,\to_1^\prime) \cap \max(X,\to_2^\prime) = c(X) = \{x\}$, a contradiction. \qed
   \end{description}    
\bigskip
%

%
\noindent \underline{\textsc{Proof of Example~\ref{EX:dem less than lib}}}.
    To prove $\dem(c) \leq 5$, let $S = \{\to_i \; : 1 \leqslant i \leqslant 5 \}$, where, for each $i$, the voter $\to_i$ only displays only the single preference $i \to_i 0$, and nothing else. 
    It is immediate to check that $S$ democratically represents $c$. The claim follows. 
    
    Next, we prove $\lib(c) \geq 10$.
    Let $\mathcal F$ be the family of all menus of size $2$ not containing the item $0$; therefore, $\vert \mathcal F \vert = {5 \choose 2} =10$. 
    We claim that each menu in $\mathcal F$ requires a different voter over $X$ in order to obtain a liberal representation of $c$: this will prove the claim. 
    Toward a contradiction, suppose that $\mathcal V$ is a liberal representation of $c$, and  let $A$ and $B$ be distinct menus in $\mathcal F$ such that $\max(A,\to) = c(A) = A$ and $\max(B,\to) = c(B) = B$ for some voter $\to$ in $\mathcal V$.
    The definition of $c$ yields $0 \in \max(A \cup \{0\}) = c(A \cup \{0\}) = A \cup \{0\}$ and $0 \in \max(B \cup \{0\})= c(B \cup \{0\}) = B \cup \{0\}$.
    It follows that $i \not\rightarrow 0$ for all $i \in A \cup B$, and so $0 \in \max(A \cup B \cup \{0\}) = c(A\cup B \cup \{0\})$.  
    However, $\vert A \cup B \cup \{0\} \vert \geq 4$ implies $0 \notin A \cup B = (A \cup B \cup \{0\})$, which is impossible.   
    \qed
    \bigskip
%

%
\noindent \underline{\textsc{Proof of Example~\ref{EX:dem = lib}}}.  
  First, we prove $\dem(c) =3$. 
  It is easy to show $\dem(c) > 2$. Indeed, if $\{\to_1,\to_2\}$ were to be a democratic representation of $c$, then we would have $x \perp_i y \perp_i z \perp_i x$ for $i = 1,2$, because $\underline{x}\underline{y}$, $\underline{x}\underline{z}$, and $\underline{y}\underline{z}$.\footnote{By $x \perp_i y$ we mean $x \not\rightarrow_i y$ and $y \not\rightarrow_i x$.}
  However, this would imply that $z$ is chosen in $\{x,y,z\}$, a contradiction.  
  To conclude the proof, we exhibit a democratic representation of $c$ having size three. 
  Let $\to_1$, $\to_2$, and $\to_3$ be the binary relations on $X$ defined as follows:\vs\vs
  $$
  x \to_1 y,z,w \;\;\hbox{ and }\;\; y \to_1 w\;,\qquad x \to_2 w \;\;\hbox{ and }\;\; y \to_2 z,w \;,\qquad w \to_3 y,z\;.\vs\vs
  $$
  Then $\{\to_1,\to_2,\to_3\}$ is a democratic representation of $c$, hence $\dem(c) =3$.
  \smallskip
  
  To prove that $\lib(c) = 3$, we first show $\lib(c) > 2$. 
  Toward a contradiction, suppose $\{\to_1,\to_2\}$ is a liberal representation of $c$. 
  Since $\underline{x}w$ and $\underline{x}w$, we have $x \to_i w$ and $y \to_i w$ for $i =1,2$. 
  Since $\underline{x}\underline{z}$, $\underline{y}\underline{z}$, and $\underline{x}\underline{y}z$, we have (without loss of generality) $x \to_1 z$ and $y \to_2 z$.  
  Since $\underline{x}\underline{z}$ and $\underline{x}zw$, we get $w \to_2 z$.
  Similarly, since $\underline{y}\underline{z}$ and $\underline{y}zw$, we get $w \to_1 z$.
  However, the join of $w \to_1 z$ and $w \to_2 z$ contradicts $\underline{z}\underline{w}$.
  We conclude that $\lib(c) > 2$.  
  (Notice that so far we have not used the selection $\underline{x}yzw$.) 
  To exhibit a liberal representation of $c$ of size three, let $\to_1$ be the strict linear order such that $x \to_1 y \to_1 z \to_1 w$, whereas $\to_2$ and $\to_3$ are the voters over $X$ defined by\vs\vs 
  $$
	x \to_2 z,w \hbox{ and } y \to_2 w \to_2 z \to_2 y \;,\quad x \to_3 w \to_3 z \,,\; y \to_3 z\,, \hbox{ and } y \leftrightarrows_3 w \;.
  \vs\vs$$
  (Note that $\to_2$ has a strict cycle, and $\to_3$ fails to be asymmetric.) 
  One can readily check that $\{\to_1,\to_2,\to_3\}$ is a liberal representation of $c$, and so $\lib(c) =3$. \qed
\bigskip
%

%
\noindent \underline{\textsc{Proof of Lemma~\ref{LEMMA:respective upper bounds}}}.
	Let $c$ be a quasi-choice over $X$.
	If Axiom$\;\alpha$ fails for $c$, then $\dem(c) = \lib(c) = \infty$, hence (i) and (ii) are verified.
	Next, suppose $c$ satisfies Axiom$\;\alpha$. 
	By Theorem~\ref{COR:full_generality}, $c$ is both democratic and liberal, and so $\dem(c)$ and $\lib(c)$ are finite. 
	In what follows, we prove (i) and (ii) by deriving a representation of a certain type (resp.\ democratic, liberal) from one of the other type (resp.\ liberal, democratic).  
	\begin{description} 
	  \item[(i):] This inequality follows from the fact 
	   that for any liberal representation of $c$, we can always create a democratic representation of $c$ by doubling the number of voters, where all new voters are neutral (i.e., they choose everything). 
	 %
	   \item[(ii):] Let $\mathcal V$ be a democratic representation of $c$ such that $\vert \mathcal V \vert = \dem(c)$.  
	   Below we construct a family\vs\vs 
	   \begin{equation*} \label{EQ:def liberal set of voters}
	   	\mathcal W_{\mathcal V} = \left\{w_{\mathcal U} : \mathcal U \subseteq \mathcal V \hbox{ and } \vert \mathcal U \vert > \textstyle \frac{\vert \mathcal V \vert}{2} \right\}\vs\vs
	   \end{equation*}
	   of ballots over $X$, which liberally represents $c$.
	   The ballot $w_{\mathcal U} \in \mathcal W_{\mathcal V}$ is such that, for each $A \in 2^X$ and $a \in X$,\vs\vs
	   \begin{equation} \label{EQ:def liberal representation (ii)}
	   	 a \in w_{\mathcal U}(A) \quad \Longleftrightarrow \quad a \in u(A) \hbox{ for all } u \in \U\,.\vs\vs
	   \end{equation}
	   To prove that $\W_{\mathcal V}$ liberally represents $c$, we show that, for each $A \in 2^X$ and $a \in X$, the equivalence\vs\vs 
	   \begin{equation} \label{EQ:relative bound (ii)}
	     a \in c(A) \quad \Longleftrightarrow \quad a \in w_\U(A) \hbox{ for some } w_\U \in \W_{\mathcal V}\vs\vs
	   \end{equation}  
	   holds. 
	   Let $A \in 2^X$ and $a \in X$.
	   If $a \in c(A)$, then, since $\mathcal V$ is a democratic representation of $c$, there is a subfamily $\U \subseteq \mathcal V$ such that $\vert \U \vert > \vert \mathcal V \vert /2$ and $a \in u(A)$ for all $u \in \U$. 
	   Now \eqref{EQ:def liberal representation (ii)} yields $a \in w_\U(A)$, which proves the forward implication in \eqref{EQ:relative bound (ii)}. 
	   We prove the reverse implication in \eqref{EQ:relative bound (ii)} by contrapositive. 
	   Suppose $a \notin c(A)$. 
	   If $a \in w_\U(A)$ for some $\U \subseteq \mathcal V$ such that $\vert \U \vert > \vert \mathcal V \vert /2$, then  \eqref{EQ:def liberal representation (ii)} gives $a \in u(A)$ for all $u \in \U$, which contradicts the fact that $\mathcal V$ is a democratic representation of $c$. 
	   Thus, there is no $\U \subseteq \mathcal V$, with $\vert \U \vert > \vert \mathcal V \vert /2$, such that $a \in w_\U(A)$.
	   Thus $\W_{\mathcal V}$ liberally represents $c$.  
	   
	   Now the bound (ii) follows easily. 
	   Indeed, letting $n := \vert \mathcal V \vert = \dem(c)$, we have: \vs\vs
	   $$
	   \vert \W_{\mathcal V} \vert \; = \; \displaystyle
	     \sum_{k=\lfloor \frac{n}{2} \rfloor +1}^n {n \choose k} \; \leqslant \; \frac{2^n}{2} \; = \; 2^{\dem(c) -1}\,,\vs\vs
	   $$
	   and so $\lib(c) \leqslant \vert \W_{\mathcal V} \vert \leqslant 2^{\,\dem(c) -1}$, as claimed. \qed
	\end{description} 
\bigskip
%

%
\noindent \underline{\textsc{Proof of Lemma~\ref{cnk-lemma}}}.
%
%
%
Let $n \geqslant 0$ and $1 \leqslant k \leqslant n+1$.
One can readily check that, for $n =0$, the choice $c_{0,1}$ is the unique choice over $X_0 = \{0\}$.
Clearly, this choice satisfies Axiom$\:\alpha$, has liberal number $\lib(c_{0,1}) = {0 \choose 0} =1$, and has democratic number $\dem(c_{0,1}) = 1$.
Thus the claim holds for $n =0$. 
  
Now suppose $n \geqslant 1$. 
Let $A \subseteq B \subseteq X_{n}$ and $x \in A \cap c_{n,k}(B)$. 
	To prove that Axiom$\:\alpha$ holds for $c_{n,k}$, we show that $x \in c_{n,k}(A)$. 
	If $x \neq 0$, then the claim readily follows from condition~\ref{cnk-lemmaa}. 
	Therefore, let $x = 0$.
	Since $0 \in c_{n,k}(B)$ by hypothesis, condition~\ref{cnk-lemmab} yields $\vert B\vert \leqslant k$, which in turn implies $\vert A \vert \leqslant k$ because $A \subseteq B$. 
	Another application of~\ref{cnk-lemmab} delivers $0 \in c_{n,k}(A)$.
	This shows that Axiom$\:\alpha$ holds for all $c_{n,k}$. 	
	
	Next, we prove the equality in \eqref{EQ:cnk-lemma}.
	Set\vs\vs 
	$$
	\mathscr{F}_{n,k} \; \coloneqq \; \big\{A \in 2^{X_{n}} : 0 \in A \text{ and } |A| = k \big\}\,.\vs\vs
	$$ 
	Plainly, $\vert \mathscr{F}_{n,k}\vert = {n \choose k-1}$. 	
	Let $\mathcal V_{n,k}=\{v_A\colon A \in \mathscr{F}_{n,k}\}$ be the family of quasi-choices on $X_n$ defined by
	$$
	v_A(B)=
	\begin{cases}
		B & B \subseteq A;\\
		B\setminus\{0\} & B \not\subseteq A.\\
	\end{cases}
	$$ 
	We show that each $v_A$ is rationalizable, and thus it is a ballot over $X_n$. 
	To that end, suppose $B \subseteq D$ and $x \in B \cap v_A(D)$.
	If $x\neq 0$, we are done.
	If $x=0$, we obtain $B\subseteq D\subseteq A$, whence $x \in v_A(B)$.	
	Next, suppose $x \in v_A(B) \cap v_A(D)$.
	If $x \neq 0$, we are done.
	Suppose $x=0$ and note that $(B\cup D) \subseteq A$.
	It follows that $0 \in v_A(B \cup D)$.
	
	Next, we prove that the family of ballots $\mathcal V_{n,k}$ liberally represents $c_{n,k}$.
	Suppose $x \in c_{n,k}(B)$ and $x \neq 0$.
	In this case, every $v_A$ is defined in a way such that $x \in v_A(B)$.
	If $x=0$ and $0 \in c_{n,k}(B)$, then we obtain $\vert B\vert\leq k$.
	Therefore we can choose a menu $A$ such that $\vert A \vert=k$ and $B\subseteq A$.
	For such a set, $0 \in v_A(B)$.
	
	Suppose there exists $A \in \mathscr{F}_{n,k}$ such that $x \in v_A(B)$.
	If $x\neq 0$, then we are readily done, because $B \setminus c_{n,k}(B)\subseteq \{0\}$.
	If $x=0$, then we have $B\subseteq A$.
	We conclude $\vert B \vert \leq \vert A \vert$, and so $0 \in c_{n,k}(B)$.
	This completes the proof that $\mathcal V_{n,k}$ liberally represents $c_{n,k}$.
	Since $\vert \mathcal V_{n,k} \vert = {n \choose k-1}$, we conclude $\lib(v_{n,k}) \leqslant {n \choose k-1}$. 
	
	To complete the proof of the equality $\lib(c_{n,k}) = {n \choose k-1}$, we show that no family $\mathcal V$ of ballots over $X_n$ having size strictly less that ${n \choose k-1}$ can liberally represent $c_{n,k}$. 
	Toward a contradiction, suppose such a family $\mathcal V$ exists. 
	Recall that $\vert \mathscr{F}_{n,k}\vert = {n \choose k-1}$, and for all $A \in \mathscr{F}_{n,k}$ we have $0 \in c_{n,k}(A)$.
	By the pigeonhole principle, there are two distinct menus $A_1,A_2 \in \mathscr{F}_{n,k}$ and a ballot $v \in \mathcal V$ such that $0 \in v(A_1) \cap v(A_2)$. 
	Apply Axiom$\:\gamma$ to obtain $0 \in v(A_1 \cup A_2)$. 
	By the liberal rationalizability of $c_{n,k}$ by $\mathcal V$, it follows that $0 \in c_{n,k}(A_1 \cup A_2)$.
	By the definition of $c_{n,k}$, we also obtain  $\vert A_1 \cup A_2 \vert \leqslant k$.   
	However, the latter fact is impossible, since $\vert A_1 \vert = \vert A_2 \vert = k$, and $A_1 \neq A_2$ implies $\vert A_1 \cup A_2 \vert > k$.
	\medskip
	
	To complete the proof of Lemma~\ref{cnk-lemma}, we show that the two inequalities in \eqref{EQ:cnk-lemma} hold. 
	%
	Define $n$ distinct ballots $\widetilde{v}_{1},\ldots,\widetilde{v}_{n}$ over $X_n$ by letting, for each $i = 1, \ldots,n$ and $A \in 2^{X_{n}}$,\vs\vs 
	\begin{equation*}
	\widetilde{v}_{i}(A) \coloneqq 
		\begin{cases}
			A \setminus \{0\} & \text{if } 0,i \in A \\
			A & \text{otherwise.}
		\end{cases}\vs\vs
	\end{equation*}
	Next, let $v^{N}$ be the neutral ballot over $X_n$ (the identity choice), and, for $n \geqslant 3$, let $v^{H}$ be the hypercritical ballot over $X_n$ (the empty quasi-choice).\footnote{For our purposes, at most $n+1$ replications of the neutral ballot will suffice. Similarly, at most $n-2$ replications of the hypercritical ballot will suffice, as long as $n \geq 3$.} 
   
    The next two claims complete the proof of the two inequalities in~\eqref{EQ:cnk-lemma}:
\smallskip

    \noindent \textsc{Claim 1:} \textit{For $\frac{n}{2} < k \leqslant n+1$, $c_{n,k}$ is democratically represented by the family}\vs\vs 
    $$
    \mathcal V_k \coloneqq \{\widetilde{v}_1,\ldots ,\widetilde{v}_n\} \cup \big\{v^N_1, \ldots , v^N_{2k-n-1} \big\},\vs\vs
    $$
    where $v^N_j = v^N$ for all $j$'s.
    \smallskip

    \noindent \textsc{Claim 2:} \textit{For $1 \leqslant k \leqslant \frac{n}{2}$, $c_{n,k}$ is democratically represented by the family}\vs\vs 
    $$
    \mathcal V_k \coloneqq \{\widetilde{v}_1,\ldots ,\widetilde{v}_n\} \cup \big\{v^H_1, \ldots, v^H_{n-2k} \big\},\vs\vs
    $$
    where $v^H_j = v^H$ for all $j$'s.
    \smallskip

    In what follows, we prove Claim~1 (Claim~2 can be proved in a similar way): this will complete the proof of Lemma~\ref{cnk-lemma}. 
    It suffices to show that $\mathcal V_k$ is such that\vs\vs 
    \begin{equation} \label{EQ:democratic choice again}
		 p \in c_{n,k}(A) \quad \Longleftrightarrow \quad \big\vert \{v \in \mathcal V_k : p \in v(A)\} \big\vert > \textstyle \displaystyle \frac{\left\vert \mathcal V_k \right\vert}{2}
	\end{equation}
	for each $A \in 2^{X_n}$ and $p \in X_n$. 
	For $p \neq 0$, the equivalence \eqref{EQ:democratic choice again} holds.  
	Suppose $p = 0$. 
	Then\vs\vs 
	$$
	\big\{ i \in \{1,\ldots n\} : 0 \in \widetilde{v}_i(A) \big\} = \big\{ i \in \{1,\ldots,n\} : i \notin A\big\} = X_n \setminus A\,,\vs\vs
	$$ 
	hence\vs\vs 
	$$
	\big\vert \big\{ i \in \{1,\ldots,n \} : 0 \in \widetilde{v}_i(A) \big\} \big\vert = n + 1 - \vert A \vert\,.\vs
	$$
	Furthermore, we have\vs\vs 
	$$
	\big\vert \big\{ i \in \{1,\ldots,2k-n-1\} : 0 \in v_{i}^{N}(A) \big\} \big\vert = 2k-n-1.\vs\vs
	$$
	Since\vs  
	\begin{equation*} 
	\big(n+1 - |A|\big) + \big( 2k-n-1 \big) > \frac{2k-1}{2} \quad \Longleftrightarrow \quad |A| \leqslant k \,,\vs
    \end{equation*}
	we conclude that the family $\mathcal V_k$ satisfies~\eqref{EQ:democratic choice again} also for $p =0$. \qed
	\bigskip\bigskip
%

%
\noindent \underline{\textsc{Proof of Theorem~\ref{THM:upper bound lib}}}. 
Let $c$ satisfy Axiom$\:\alpha$ over a grand set $X$ of size $n \geqslant 2$.
For simplicity, in the following we identify $X$ with $\{1,\ldots,n\}$.\vs\vs 

\begin{description}
	\item[(Upper bound)] For each $p \in X$, denote by $\mathscr{A}_{p}$ the collection of the $\subseteq$-maximal sets $A \subseteq X$ such that $p \in c(A)$. 
	Let $\mathcal V^{*}$ be any collection of ballots over $X$ satisfying the following properties:\vs\vs
\begin{enumerate}[label=(\roman*)]
\item for every ballot $v \in \mathcal V^{*}$, there exist $C_{1} \in \mathscr{A}_{1},\ldots,C_{n} \in \mathscr{A}_{n}$ such that\vs\vs
\[
v(B) = \{ p \in X : B \subseteq C_{p} \},\quad \hbox{for $B \in 2^{X}$};\vs\vs
\]
%
%
\item for every item $p \in X$,\vs\vs
\begin{align*}
\Big\{ \bigcup_{\overset{B \in 2^{X}}{p \in v(B)}} B : v \in \mathcal V^{*} \Big\} = \mathscr{A}_p;\vs\vs
\end{align*}
%
%
\item\label{triple_i} $|\mathcal V^{*}| \leqslant \max_{p \in X} |\mathscr{A}_{p}|$.\vs
\end{enumerate}
It is not difficult to check that $\mathcal V^{*}$ liberally represents $c$. 
To complete the proof, we show $|\mathcal V^{*}| \leqslant {n -1 \choose \lfloor \frac{n-1}{2} \rfloor}$. 
For $p \in X$, let\vs 
$$
\mathscr{A}_{p}' \coloneqq \{ A \setminus \{p\} : A \in \mathscr{A}_{p}\},\vs
$$
so that $|\mathscr{A}_{p}'| = |\mathscr{A}_{p}|$ holds.
Plainly, each  $\mathscr{A}_{p}'$ is a Sperner family over $X \setminus \{p\}$, as its members are mutually $\subseteq$-incomparable.,
Therefore, by Sperner's theorem \citep{Sperner28,Lubell66}, we have $|\mathscr{A}_{p}'| \leqslant {n -1 \choose \lfloor \frac{n-1}{2} \rfloor}$, and so by \ref{triple_i} we get $|\mathcal V^{*}| \leqslant {n -1 \choose \lfloor \frac{n-1}{2} \rfloor}$. 
We conclude that $\lib(c) \,\leqslant\, {n -1 \choose \lfloor \frac{n-1}{2} \rfloor}$ holds.

\item[(Tightness)]
Take the liberal choice $c = c_{n-1,\lfloor \frac{n-1}{2}\rfloor +1}$ over $X_{n-1}$ defined in Lemma~\ref{cnk-lemma}, and observe that $\vert X_{n-1} \vert =n$ and $\lib(c) = {n-1 \choose \lfloor \frac{n-1}{2} \rfloor}$. \qed
\end{description}

\end{document}